\documentclass[aps,prb,twocolumn,superscriptaddress]{revtex4-1}

\usepackage{graphicx}

\begin{document}


\title{Polarization-preserving confocal microscope for optical experiments\\
in a dilution refrigerator with high magnetic field}
\author{Maksym~Sladkov}

\author{M.~P.~Bakker}

\author{A.~U.~Chaubal}

\affiliation{Zernike Institute for Advanced Materials,
University of Groningen, NL-9747AG Groningen, The
Netherlands}

\author{D.~Reuter}
\author{A.~D.~Wieck}
\affiliation{Angewandte Festk\"{o}rperphysik,
Ruhr-Universit\"{a}t Bochum, D-44780 Bochum, Germany}

\author{C.~H.~van~der~Wal}
\affiliation{Zernike Institute for Advanced Materials,
University of Groningen, NL-9747AG Groningen, The
Netherlands}

\date{\today}

\begin{abstract}
We present the design and operation of a fiber-based
cryogenic confocal microscope. It is designed as a compact
cold-finger that fits inside the bore of a superconducting
magnet, and which is a modular unit that can be easily
swapped between use in a dilution refrigerator and other
cryostats. We aimed at application in quantum optical
experiments with electron spins in semiconductors and the
design has been optimized for driving with, and detection
of optical fields with well-defined polarizations. This was
implemented with optical access via a polarization
maintaining fiber together with Voigt geometry at the cold
finger, which circumvents Faraday rotations in the optical
components in high magnetic fields. Our unit is versatile
for use in experiments that measure photoluminescence,
reflection, or transmission, as we demonstrate with a
quantum optical experiment with an ensemble of donor-bound
electrons in a thin GaAs film.
\end{abstract}

\maketitle

\section{Introduction}
Fundamental research on quantum coherence in solid state is
currently strongly driven by the goal to implement quantum
information tasks
\cite{Awschalom2002,Zoller2005,Southwell2008}. Electron and
hole spins in semiconductors are here of special interest
since these material systems give access to realizing
compact devices where quantum correlations between
coherence of spins and optical signal fields can be
established. However, optical manipulation and detection of
spins is a challenging task since experiments demand a
combination of conflicting requirements such as cryogenic
temperatures, high optical intensities, high magnetic
fields, and precise control of the optical polarizations.

In recent years research in this direction nevertheless
developed into a very active field that showed many
impressive results, in particular with localized spins in
quantum dots or spins bound to donor sites
\cite{Fu2005,Atature2007,Fu2008,Press2008,Berezovsky2008,Greilich2009,Xu2009,Brunner2009,Latta2009a}.
Most of these experiments were carried out at temperatures
of a few Kelvin. Lower temperatures were till now typically
not useful since spin states can be prepared via optical
pumping, and the coherence of localized spins is in all
III-V and most II-VI semiconductors limited by hyperfine
coupling to fluctuating nuclear spins. We anticipate,
however, that dephasing by nuclear spins can be removed
with controlled dynamical nuclear polarization effects
\cite{Xu2009,Latta2009a} or with spin echo techniques
\cite{Clark2009}. If such techniques are fully exploited,
optical experiments at milliKelvin temperatures become of
interest, since freezing out the phonons at energies of the
Zeeman splitting of electron or hole spins will be
important for exploring the ultimate limit of controlling
spin coherence in solid state. We report here the design
and operation of a fiber-based confocal microscope in a
dilution refrigerator that is suited for this research.
This instrument also has relevance for related quantum
optical studies with spin coherence of quantum Hall states
\cite{VanderWal2009,Imamoglu2000}, or optical studies of
the Kondo effect \cite{Tureci2009,Kleemans2010,Latta2010}.

When designing optical experiments in a dilution
refrigerator one has to deal with a number of restrictions.
(\textit{i}) The sample space is typically inside a series
of heat shields that forbid free-space optical access to
the sample. The sample volume can be reached with optical
fibers, but this requires a method to couple light from the
fiber to the sample. (\textit{ii}) The limited sample
volume requires compact solutions which can withstand
cryogenic temperatures and high magnetic fields.
(\textit{iii}) Using high optical intensities is in
conflict with the need to maintain low power dissipation.
High intensity at low optical power can obviously be
realized with tight focussing, but this requires a
microscope with good performance in the sample volume.
(\textit{iv}) If the experiments require well-defined
polarizations, one needs to deal with the fact that in high
fields most optical components will cause a substantial
Faraday rotation of the polarization.

To address all these constraints we build on an earlier
design of a fiber-based confocal microscope for use
at 4.2 Kelvin \cite{Hogele2008}. Our key innovations are that we
circumvent Faraday rotations by using a polarization
maintaining fiber and by using a compact cryogenic
microscope with all propagation through free-space elements
in a direction that is orthogonal to the magnetic field. In
addition, we make the approach suited for application at
milliKelvin temperatures and we report how we deal with
stronger constraints for the heat load from various
functions, as for example the wiring to piezo motors that
drive the microscope focussing.

We demonstrate application of the microscope on an ensemble
of donor-bound electrons in a thin GaAs film. We performed
optical spectroscopy of spin-selective transitions of the
donor-bound electron states to donor-bound trion states,
both via photoluminescence and in direct transmission
experiments. We also directly demonstrate a quantum optical
effect, known as electromagnetically induced transparency,
with this material system.

\section{Microscope design and operation}

\subsection{Modular compact microscope unit}

We designed the confocal microscope as a compact
cold-finger that fits inside the bore of a superconducting
magnet, and which is a modular unit that can be easily
swapped between use in a dilution refrigerator (Leiden
Cryogenics DRS1000) and a helium bath cryostat. Both
systems are equipped with a superconducting magnet
(Cryogenics Ltd.), with a bore that yields a cylindrical
sample space of 60 mm diameter (for the dilution
refrigerator this concerns a 78 mm bore with a series of
heat shields around the sample volume). In both cryostats
the magnetic field is applied along the vertical direction
(defined as $z$-direction).

\begin{figure}
  \includegraphics[width=1.0\linewidth]{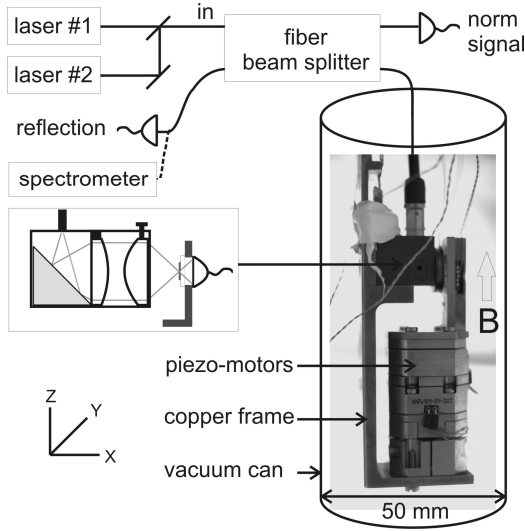}\\
  \caption{(color online) Schematics of the experimental setup.
  Excitation light of two tunable lasers is coupled
  into a polarization-preserving fiber-based beam
  splitter (port IN) and one of the outputs is
  connected to the fiber that runs to the microscope.
  This fiber delivers excitation light to the
  sample, which is mounted on an $xyz$-stack of piezo-motors.
  The sample position can be tuned to be in or out
  of the focal spot of the two-lens microscope.
  The microscope is mounted in a tube, which is
  vacuum pumped and immersed in a Helium
  bath (4.2~K) or used in a dilution
  refrigerator. A superconducting
  coil provides magnetic fields up to 9~T. A
  silicon \textit{pin}-photodetector is positioned
  right behind the sample for detection of the
  optical transmission. Both the sample and the detector are
  mounted on a $\Gamma$-shaped sample holder. The
  second output of the beam splitter is coupled to
  a photodetector for monitoring the optical powers.
  Signals that come from reflection on the sample, as well
  as emission by the sample, retraces the optical
  path through the fiber.
  After passing the beam splitter is can be
  diverted to a regular photodetector, or
  to a spectrometer.
  Inset: microscope components mounted on the
  copper frame that forms the cold finger.}\label{fig:microscope}
\end{figure}

\subsection{Polarization maintaining fiber}

For delivering light into the sample volume we use a
single-mode polarization-maintaining PANDA-type fiber
\cite{Fiber} (PMF) with $NA=0.13$. Its mode-field diameter
is $5~\rm{\mu m}$ and the cut-off wavelength is
$700~\rm{nm}$. Operation of the PMF is based on built-in
stress that induces a linear birefringence (different index
of refraction, $n_V$ and $n_H$) for two TEM propagation
modes with orthogonal linear polarizations. Linear light
coupled into one of these two modes does not couple into
the other mode during propagation. We apply this for
delivering fields with polarization parallel (defined as V
polarization) and orthogonal (defined as H polarization) to
the magnetic field.

Using a PMF has the advantage that we can deliver
well-defined polarizations that remain pure in the field
range -9~T to +9~T. This removes the need for
\textit{in-situ} polarization analysis for calibrating and
pre-compensating Faraday rotations and other effects that
influence the polarizations \cite{Hogele2008}. In addition,
the polarizations remain pure when the fiber is subject to
stress from bending and thermal gradients, and the
polarization purity is thereby also more resilient against
mechanical vibrations.

To evaluate whether the PMF will indeed suppress all
Faraday rotations in fields up to $\pm$9~T we need to
compare the stress induced birefringence to the Faraday
effect. This can be quantified by comparing the associated
beat length and rotation length, respectively. If linearly
polarized light is coupled into the PMF but not along one
of the TEM eigenmodes, its polarization will undergo
periodic unitary transformation due to the difference
between $n_V$ and $n_H$. The period of transformation is
the linear birefringence length $L_{LB}$ (the smaller
$L_{LB}$, the better the polarization maintaining
properties of the fiber). For the fiber in use \cite{Fiber}
$L_{LB}=2.4~\rm{mm}$. The Faraday effect occurs when an
optical fiber is a subject to magnetic field along its
optical axis. It then shows circular birefringence and its
strength is characterized by the Verdet constant. This
yields a circular birefringence rotation length $L_{CB}=2
\pi /(V \cdot B) \approx 4~\rm{cm}$ for a magnetic field
$B=9~\rm{T}$ and with the Verdet constant of the fiber
\cite{Fiber} $V=6~\rm{rad~T^{-1} m^{-1}}$. Our fiber thus
has $L_{LB} < L_{CB}$ and we therefore expect that linear
polarization of light coupled into one of the fiber TEM
eigenmodes will not be affected by magnetic field. We do
check that this is indeed the case by using polarization
selective optical transitions of donor-bound electrons as a
polarization probe (further discussed in
Sec.~\ref{sec:characterization}).

The single-mode nature of the PMF is useful for experiments
where two co-propagating fields should drive the same
system. The beam overlap for these two fields will then be
ideal in the sample volume.

\subsection{Confocal microscope in Voigt geometry}

To focus light from the fiber in the sample volume we use a
compact home-built microscope objective, based on two
aspheric lenses. The approach is similar to the system used
by A. H\"{o}gele \cite{Hogele2008} \textit{et al.}, but we
incorporated changes that avoid circular birefringence that
occurs when light propagates through lens material in high
fields. This can rotate linear polarizations by several
degrees in $1~\rm{mm}$ lens material in a $9~\rm{T}$
magnetic field \cite{Munin1992} (typical values for the
Verdet constant of materials \cite{Valev2008,Robinson1964}
as BK7 and Corning glass are 1--10~${\rm rad~T^{-1}
m^{-1}}$). Before light is collimated and focused its
propagation direction is diverted $90^o$ by means of a
surface mirror (dielectric prism, Thorlabs MRA05-E02) after
which its $\vec{k}$-vector is orthogonal to the direction
of the magnetic field and this eliminates the magnetic
field induced rotation of the polarization.

Another advantage of this approach is that it gives access
to optical experiments in Voigt geometry (light propagation
direction orthogonal to the magnetic field direction). In
this geometry, light with V polarization can drive atomic
$\pi$-transitions (no change in angular momentum), and
light with H polarization can drive atomic
$\sigma$-transitions (a change in angular momentum of $\pm
\hbar$). Thus, this yields that for the typical optical
selection rules of electron spins in semiconductors,
transitions that start from the electron spin-up or
spin-down state can be addressed selectively with two
orthogonal linear polarizations. Another advantage is that
one of these two fields can be easily blocked by
polarization filtering before the two co-propagating modes
reach a detector (as required for several resonant-Raman
control schemes \cite{Duan2001}). Such a filter with high
extinction ratio can be realized before any propagation
through glass material as a metal nanowire-grid polarizer
on a transparent substrate \cite{Wang2006}.

The first lens of the microscope objective collimates the
light from the fiber. We use an aspheric lens (Thorlabs
350430) with focal length $f=5.0~\rm{mm}$, numerical
aperture $NA=0.15$ and clear aperture $1.5~\rm{mm}$. After
collimating, light is focused by a second aspheric lens
(Thorlabs 350140) with focal distance $f=1.45~\rm{mm}$,
$NA=0.55~\rm{mm}$ and clear aperture $1.6~\rm{mm}$. Since
we do not have perfect $NA$ matching between the fiber and
the collimating lens, the numerical aperture of the full
objective system will be less than that of the focusing
lens, resulting in $NA_{obj}=0.45$. Using this we can
estimate the theoretical value for the spot diameter to be
$D_{spot}=1.42 \cdot \lambda$.

The prism mirror and the collimating lens are firmly glued
into a DuPont Vespel-SP1 housing. This material was chosen
(instead of copper for example) since its thermal expansion
is smaller, and also lies closer to that of the lenses. The
focusing lens is fixed inside a separate Vespel-SP1 frame
and then fitted into the housing of the objective. This
allows for removing the focusing lens from the objective
while aligning the fiber with respect to the collimating
lens. In addition, the focussing lens has a certain
positioning freedom (along the propagation direction, about
$1~\rm{mm}$). This can be used to fit in a plastic
thin-film $\lambda /4$ retardation plate, which can be used
if one wishes to transform the linear polarizations into
circular polarizations.

\begin{figure}
  \includegraphics[width=40mm]{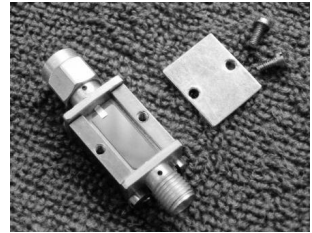}\\
  \caption{Photo of a home-built heat-sink for
  use in cryogenic coaxial lines,
  with SMA connectors and a gold-plated sapphire
  substrate in a copper housing unit.}\label{fig:smaheatsink}
\end{figure}

\subsection{Focussing mechanics and wiring to the piezo motors}

For positioning the sample in the focus of the objective we
use a three-axes $xyz$-translation stage on the cold finger
in the form of slip-stick piezo-motors (Attocube, model
ANP101). This gives a travel range of $5~\rm{mm}$ in all
three directions. Both the piezo stack and the objective
are fixed on a home-built copper frame
(Fig.~\ref{fig:microscope}).

Applying such piezo motors in a dilution refrigerator is
not straightforward since the wires to the motors should
have a resistance of $5~\rm{\Omega}$ at most. In the
dilution refrigerator the length of the wires exceeds a few
meters, such that wires with low resistivity must be used.
This, however, creates an unacceptable heat load on the
mixing chamber with most types of wiring. We circumvented
this problem by using coaxial lines with low DC resistance
values. This works very well since the need for the low
resistance values is partly driven by the need for high
bandwidth for getting triangular control pulses
($\sim$100~V) with a fast edge to the piezo-motors without
much smoothing.

We implemented this by installing a set of 7 of
multi-purpose coaxial lines to the mixing chamber
(Micro-Coax, semi-rigid model UT 85-B-SS, 60~GHz bandwidth,
stainless steel outer conductor, silver-plated BeCu inner
conductor, $\sim$0.5~$\Omega {\rm m}^{-1}$ DC resistance).
These have an excellent trade-off between low electrical
resistance and low thermal conductance at low temperatures.
The piezo-motors use 3 of these coaxial lines.

Realizing a low heat load on the mixing chamber still
requires to heat-sink the inner conductor of the coaxial
lines at several stages between room temperature and base
temperature. We use home-built heat sinks at 4.2~K, the
1K-pot ($\sim$1.5~K), the still ($\sim$600~mK), the cold
plate ($\sim$100~mK) and at the mixing chamber. The heat
sinks are made with sapphire substrates in a copper housing
unit (Fig.~\ref{fig:smaheatsink}), since sapphire combines
high electrical resistance with good thermal conductivity
at low temperatures. We gold-plated one of the surfaces of
the sapphire. The non-plated sapphire surface is then glued
onto a copper plane in the housing unit with thin stycast.
Two SMA connectors are then mounted on this unit, and we
solder the inner conductors on opposite sides of the gold
plated sapphire surface. Our units show more than
sufficient bandwidth (a few MHz) for the piezo-motor
application, but this can be increased by engineering the
sapphire plates as microwave strip lines. After installing
these coaxial lines we did not observe an effect on the
cooling power of the dilution refrigerator (base
temperature well below 20~mK), and we achieved driving of
the piezo-motors via these coaxial lines without any
problems.

\subsection{Optical detection and cooling down procedure}

For optical detection in transmission experiments we use a
photodiode (Hamamatsu, $pin$, 5106) on the $\Gamma$-shaped
copper sample holder that is mounted on the piezo-motor
stack (Fig.~\ref{fig:microscope}). The sample can be
mounted in front of the photodiode. The detection of
signals from reflection and photoluminescence experiments
is discussed below.

The use of the photodiode on the cold finger shows in
practice the risk that it breaks while cooling down the
system. Most likely, this is because thermal shrinking
causes a crack in the plastic laminate of the diode that
also breaks the semiconductor chip underneath. To avoid
this, we cool down with He contact gas at a pressure which
does not exceed $10^{-2}~\rm{mbar}$. This enforces the
system to cool very slowly (in about 6 hours). With this
approach, our photodiodes survive in 9 out of 10 cases.

\subsection{Focal plane positioning and spot size}

\begin{figure}
  \includegraphics[width=1.0\linewidth]{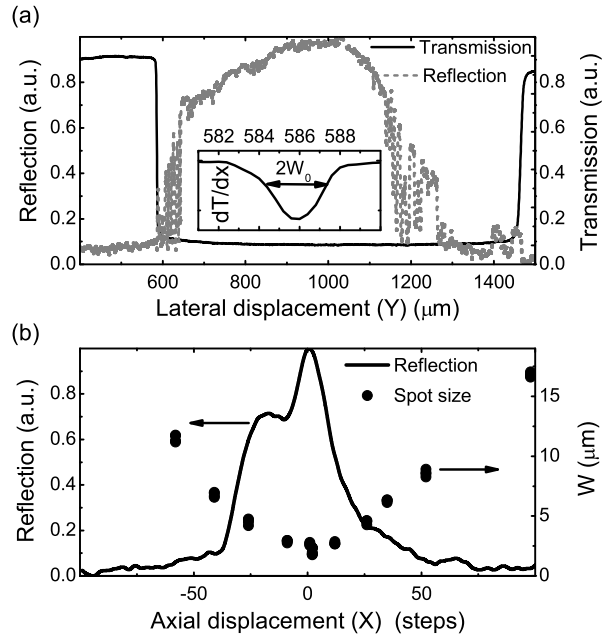}\\
  \caption{(a) Reflected and transmitted signal as a
  function of the lateral position ($y$-axis) of the
  sample. The spot size $W$ of the beam on the sample surface
  (waist $W_0$ when in focus)
  is determined with the
  knife-edge technique (inset). (b) Reflected signal (solid line)
  and the spot size $W$ (dots)
  as a function of the axial displacement ($x$-axis) of the sample.}
  \label{fig:focalplane}
\end{figure}

Since we do not have direct free optical access to the
system, it is necessary to design a procedure for
positioning and focussing that rely on using the
transmitted and reflected signal only. For positioning in
the plane orthogonal to the optical axis, we use the
trivial approach with high-contrast markers on the sample
that can be detected in the transmission signal. For
positioning the sample plane in the focal point of the
objective (along $x$-axis), we use the fact that the output
of the fiber and the objective unit together constitute a
confocal microscope. We use this in a procedure with a
fiber-coupled diode laser (Thorlabs LPS-785-FC) operated
below the lasing threshold (incoherent light for avoiding
interferences in the back-reflected signal). The light is
delivered into the sample volume via the optical fiber,
with a fiber-based beam splitter (OZ optics) in its path
(Fig.~\ref{fig:microscope}). This allows for measuring the
reflected signal while the position of the sample surface
is scanned along the $x$-axis through the focus of the
objective.

A typical focal plane scan is shown in
Fig.~\ref{fig:focalplane}(b). As expected, the reflected
signal reaches a maximum when the sample plane is
positioned exactly in the focus of the objective and drops
smoothly while going out of focus. In contrast to the
previously reported microscopes \cite{Hogele2008} we
observe not a single Lorentzian reflection profile but a
superposition of two Lorentzians, which indicates a small
misalignment within the confocal microscope. This is most
likely due to a non-uniaxial arrangement of the lenses with
respect to the optical axis of the fiber that results from
a small misalignment of the prism mirror. We did not
correct this for the data presented here since we have the
prism mirror firmly glued in the objective housing, and it
did not compromise the experiments we present below.

To determine the spot size we use the knife-edge technique,
where we fix the axial position (along the $x$-axis) of the
microscope at different locations on the reflection curve
(Fig.~\ref{fig:focalplane}(b)) and perform a lateral scan
($y$-direction) across the sample. A typical scan, taken
with the sample plane in focus, is shown in the
Fig.~\ref{fig:focalplane}(a). The solid black curve is the
transmitted signal and the dashed gray curve is the
reflected signal. As expected, a high signal level in
reflection corresponds to low signal level in transmission
since we use here an $n$-GaAs sample that is opaque for the
used wavelength, on a transparent substrate. Light that is
reflected from the sample surface is efficiently collected
back into the fiber because of the confocal geometry. The
reflection profile is also influenced by the morphology of
the sample surface, which is not uniform due to fabrication
imperfections. It is, however, always possible to find a
spot on the sample with a clean reflection profile.

Since we know the size of the sample we can calculate the
lateral ($y$-axis) step size of the piezo-motors (which
depends on temperature and mechanical load). From the slope
of the transmission signal when scanning across a sharp
edge we can determine the spot size of the beam (inset of
the Fig.~\ref{fig:focalplane}(a)). We define the spot size
on the surface of the sample as the radius $W$ of a
Gaussian beam (waist $W_0$ when in focus). The black dots
in the Fig.~\ref{fig:focalplane}(b) are results for $W$
after fitting the knife-edge profile with a Gaussian
function.

The measured beam waist was found to be $W_0=2.3~\rm{\mu
m}$, which is almost four times larger than our theoretical
estimate for the diffraction limited spot size. This is
related to the small misalignment, since the aspheric
lenses are highly optimized for realizing a small focus
exactly on the optical access, and with correcting the
alignment this type of objective at low temperatures can
yield a spot size that is well within a factor 2 from the
diffraction limit \cite{Hogele2008}. The larger spot was
compromising the efficiency of collecting light in our
photoluminescence experiments, but it did not compromise
our transmission experiments (further discussed below). In
the transmission experiments we also work at least a small
amount out of focus in order to eliminate interference
effects that occur when working in focus. These then result
from the Fabry-Perot cavity that is formed between the
sample surface and the facet of the fiber output.

\section{Characterization of performance \label{sec:characterization}}

A characterization of the focussing protocol and the spot
size was already included in the previous section. In this
section we focus on the polarization purity and achieving a
low heat load and good heat-sinking on the cold finger.

\subsection{Polarization purity \label{subsec:polarizationpurity}}

\begin{figure}
  \includegraphics[width=1.0\linewidth]{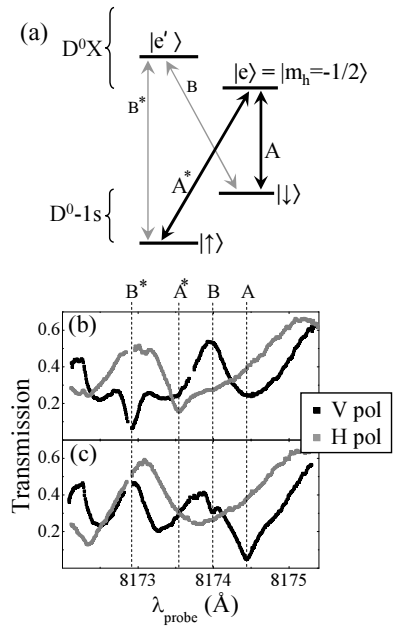}\\
  \caption{(a) Energy level diagram and optical
  transitions for the $D^0$-$D^0 X$ system in GaAs.
  (b) Pump-assisted spectroscopy with
  pumping V-polarized light at the $A$ transition
  (at $8174.45~\rm{\AA}$).
  This results in enhanced absorption for H-polarized light
  at the $A^*$ transition (8173.55~\rm{\AA}).
  (c) Complementary to the observation of (b),
  pumping with H-polarized light at the $A^*$
  transitions results in
  enhanced absorption for V-polarized light
  at the $A$ transition.
  Similar cross-pumping effects are observed for the
  nearby $B$ and $B^*$ transitions.
  Data taken at $B=8~{\rm T}$.}
  \label{fig:PumpAsSpec}
\end{figure}

In order to characterize the polarization preserving
properties of the setup we used the optical transitions of
donor-bound electrons ($D^0$ system) to a donor-bound trion
state (the lowest level of the $D^0X$ complex) in a strong
magnetic field. Our sample was a thin GaAs epilayer with Si
doping at very low concentration (further discussed below).
The relevant level scheme with transitions labeled $A$ and
$A^*$ is presented in Fig.~\ref{fig:PumpAsSpec}(a). The
states $\mid\uparrow\rangle$ and $\mid\downarrow\rangle$
are the spin-up and spin-down state of the electron in the
$D^0$ system, which is localized at the donor site in a
Hydrogen-like $1s$ orbital. These two states are
Zeeman-split by the applied magnetic field. The optical
transitions $A$ and $A^*$ are to the lowest energy level of
the $D^0 X$ system (state $|e\rangle$,  well separated from
the next level $|e ' \rangle $), which has two electrons in
a singlet state and a spin-down hole with $m_{h}=-\frac12$
localized at the donor site \cite{Clark2009}. The optical
selection rules of this system have been characterized very
well \cite{Karasyuk1994} and show a strong polarization
dependence. For the $D^0$ system, H-polarized light couples
to the $A^*$ transition (with a change in angular momentum
of $\hbar$) but not to the $A$ transition. In contrast,
V-polarized light couples to the $A$ transition (and not to
$A^*$), since this is a transition without a change in
angular momentum.

For performing this test we took scanning-probe
transmission spectra with tunable CW Ti:sapphire lasers
(Coherent MBR-110, linewidth below 1~MHz) around the $D^0
X$ resonances (Fig.~\ref{fig:PumpAsSpec}(b,c)). These
spectra are results of pump-assisted transmission
spectroscopy. This approach is needed for avoiding
bleaching of transitions due to optical pumping by the
probe, and it is also useful for identifying whether
spectral lines are from transitions that start from
$\mid\uparrow\rangle$ or from $\mid\downarrow\rangle$. For
further explanation it is best to focus on a typical result
(Fig.~\ref{fig:PumpAsSpec}(c)): Here, we fixed the pump
laser on the frequency of the $A^*$ transition with H
polarization, while we scan the probe laser frequency with
V polarization and study its transmission. We then observe
that the absorption of the $A$ transition is strongly
enhanced by the pump (compare to absorption by $A$ in
Fig.~\ref{fig:PumpAsSpec}(b)). These results appear with
respect to a background signal with slower modulation of
the transmission that is due to a Fabry-Perot effect in the
sample (further discussed below). In these experiments we
use a chopper in the probe beam and lock-in techniques for
separating the detected signal from the pump and probe
beam.

The strong polarization dependence of the absorption lines
$A$ and $A^*$ in the transmission spectra in
Fig.~\ref{fig:PumpAsSpec}(b,c) demonstrates that the linear
polarizations H and V are indeed well preserved in our
setup. We performed such experiments with magnetic fields
in the range from $B=5$ to 9~T, and the effective
polarization selectivity did not show a dependence on $B$.
We analyzed that in our experiments the characterization of
the polarization purity is in fact limited by the accuracy
of the polarization preparation on the optical table, and
the alignment of the H and V mode of the PMF fiber output
with respect to the direction of the magnetic field.
Characterizing this with all instrumentation at room
temperature showed that the error from coupling light
purely into one of the eigenmodes of the PMF is at the
level of 1 part in 100. In our experiment this was
sufficiently low, evidenced by the fact that an attempt to
pump transition $A$ with H-polarized light did not induce
any changes in the transmission spectrum from such
pump-assisted spectroscopy.

For characterizing the polarizations we could not implement
the approach of A. H\"{o}gele \textit{et al.}
\cite{Hogele2008} with cryogenic polarizing beam splitters
near the sample volume. Our sample space is too small, and
it would also be difficult to deal with Faraday rotations
in the beam splitter cube itself.

\subsection{Heat load and heat-sinking of optical power}

With our unit installed in the dilution refrigerator, we
could cool down to milliKelvin temperatures without any
problems, and running the optical experiments with typical
conditions did not show excessive heating. In particular,
in the next section we discuss experiments in which we
drive the $A$ transition with optical Rabi frequencies
$\Omega_c$ up to $2\pi \cdot 2$~GHz. We performed this with
a spot size as large as 200~${\rm \mu m^2}$, where this
driving corresponds to an optical heat load of 20~${\rm \mu
W}$. This is well below the cooling power of our dilution
refrigerator at 100~mK.

Obviously, the semiconductor material in the focus of the
optical fields can be at a much higher temperature than the
mixing chamber material. We indeed found that it is crucial
to thoroughly heat-sink our $n$-GaAs epilayers. The only
experiments where we could avoid this heating used $n$-GaAs
epilayers that were directly attached to a sapphire
substrate with binding by van der Waals forces, and a good
thermal contact between the sapphire and the cold finger.
These samples were prepared by first using an epitaxial
lift-off technique \cite{Yablonovitch1987} for removing the
GaAs epilayers from the original GaAs wafer. This was
followed by transferring the epilayer to a wedged sapphire
substrate.

While we thus successfully operated this unit at sub-Kelvin
temperatures, the experimental data did not show features
that differed from the 4.2~K data. For the experiments on
$D^0$ system that we present below, this could be expected.
The optical transitions at 4.2~K show an inhomogeneous
linewidth of 6~GHz, which is not expected to narrow at
lower temperatures. Also, the electron spin dephasing time
for the $D^0$ system is limited by hyperfine coupling to
fluctuating nuclear spins, and this mechanism is
temperature independent in the range 10~mK to 4.2~K. Our
attempts to suppress these fluctuations via dynamical
nuclear polarization (DNP) only showed a very small
improvement of the dephasing time till now, since the
optically induced DNP effects appeared to be very weak in
our experiments \cite{Sladkov2010}. For the discussion of
applications of our system we therefore focus on data taken
at 4.2~K.

\section{Application: Spectroscopy and EIT with $n$-GaAs}

In order to demonstrate the versatility of our setup we
performed an optical study of the coherent properties of an
ensemble of $D^0$ systems in GaAs. This system was already
introduced in Sec.~\ref{subsec:polarizationpurity}. We
present here data from epitaxially grown $10~\rm{\mu m}$
films of GaAs with Si doping at $n_{Si}=3\times10^{13}~{\rm
cm^{-3}}$. At such low concentrations the wavefunction of
the neighboring donors do not overlap, which yields an
ensemble of non-interacting $D^0$ systems.

To study the coherent properties of such ensembles we first
find the spectral position of the relevant optical
transitions \cite{Bogardus1968} with photoluminescence and
transmission spectroscopy. We then resolve the fine spectra
of the Zeeman-split levels \cite{Karasyuk1994} with the
pump-assisted transmission spectroscopy. After identifying
the $D^0$-$D^0X$ system in this manner, we could
demonstrate electromagnetically induced transparency
\cite{Fleischhauer2005a} (EIT) with this medium, which is a
quantum optical effect that uses the $D^0$ spin coherence.

\subsection{Photoluminescence}

We performed photoluminescence experiments to identify the
spectral region where emission by the donor-bound excitons
occurs. We brought excitation light (wavelength
$\lambda=8050~\AA$) to the sample with a  fiber-based beam
splitter in the optical path (Fig.~\ref{fig:microscope}).
The reflection channel of the beam splitter is coupled to a
PI Acton spectrometer, equipped with a nitrogen cooled CCD
camera. The light at the excitation wavelength was
suppressed by passing the reflected signal through a
bandpass filter $\lambda_c=8200~{\rm\AA},~\Delta
\lambda=100~{\rm\AA}$). The sample surface was positioned
in the focus of the confocal microscope in order to
maximize efficiency of the luminescence collection.

\begin{figure}
  \includegraphics[width=1.0\linewidth]{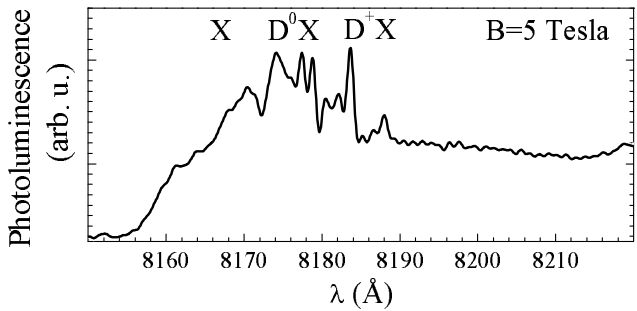}\\
  \caption{Photoluminescence spectrum of low-doped
  $n$-GaAs,
  showing
  luminescence by free excitons ($X$), excitons bound to
  neutral donor sites ($D^0 X$),
  and excitons bound to ionized donor sites in the
  sample's depletion layer ($D^+X$).
  Data taken at $B=5~\rm{T}$,
  the resolution of the spectrometer is $\sim 0.2~\rm{\AA}$.}
  \label{fig:PLspec}
\end{figure}

A typical photoluminescence spectrum taken at $B=5~\rm{T}$
is shown in Fig.~\ref{fig:PLspec}. The spectrum is
dominated by three structured peaks, which result from
emission by free excitons ($X$), excitons bound to neutral
donor sites ($D^0 X$), and excitons bound to ionized donor
sites in the sample's depletion layer near the surface
($D^+X$). The fine structure due to the Zeeman splitting of
the electron and hole spins of the ($D^0X$) and ($D^+X$)
bound excitons is observed, but does not provide sufficient
information for identifying all the transitions due to the
highly unequal oscillator strengths \cite{Fu2005}.

\subsection{Transmission spectroscopy}

\begin{figure}
  \includegraphics[width=1.0\linewidth]{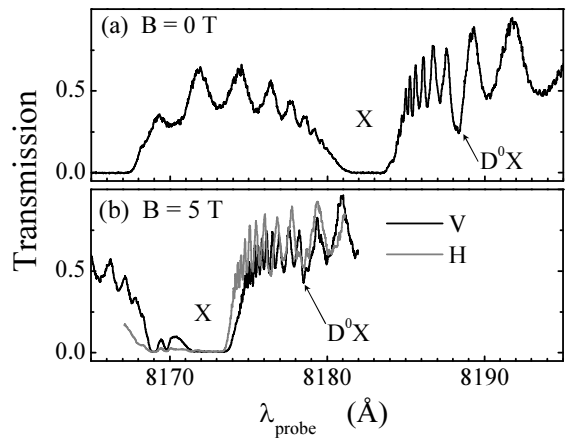}\\
  \caption{Scanning-probe transmission spectra
  taken at $B=0~\rm{T}$ (a) and
  $B=5~\rm{T}$ (b). Traces were recorded with
  linear H or V polarization for the probe light
  (giving identical results at 0~T).
  For performing these experiments the
  microscope was defocussed to a spot diameter of about
  $16~{\rm \mu m}$.
  The strong absorption due to free excitons ($X$)
  and much weaker features due to donor-bound excitons ($D^0X$)
  are labeled.
  The data at 5~T shows a diamagnetic shift of about $10~{\rm \AA}$
  with respect to the data at 0~T.}
  \label{fig:SingLasSpec}
\end{figure}

While photoluminescence is useful for initial
characterization of a material and for finding the main
spectral features, a more detailed characterization of the
$D^0$ systems requires transmission spectroscopy with
tunable CW lasers. A few results of this approach were
already presented in Fig.~\ref{fig:PumpAsSpec} and
discussed in Sec.~\ref{subsec:polarizationpurity}. Before
performing such pump-assisted transmission spectroscopy on
the $D^0$ systems, we first take scans with a single laser
over a much larger wavelength region for finding the $D^0$
lines. Figure~\ref{fig:SingLasSpec}(a,b) show such
transmission spectra taken with H and V-polarized probe
light. These results show a strong free-exciton absorption
band (labeled $X$), and weaker $D^0X$ resonances. The
oscillating background that is superimposed on the
transmission spectra is due to a Fabry-Perot effect in the
$10~\rm{\mu m}$ GaAs film, and its chirped wavelength
dependence is due to the wavelength-dependent refractive
index associated with the strong free exciton absorption.
After locating the $D^0$ lines, we zoom in on this region
for identifying the $A$ and $A^*$ transitions with the
pump-assisted spectroscopy, as presented in
Fig.~\ref{fig:PumpAsSpec}(b,c).

\subsection{Electromagnetically
induced transparency with donor-bound electrons}

We identified the $A$ and $A^*$ transitions because it was
our goal to investigate whether these could be used for
implementing electromagnetically induced transparency (EIT)
with electron spin coherence in a semiconductor. EIT is the
phenomenon that an absorbing optical transition becomes
transparent because destructive quantum interference with
another driven optical transition prohibits populating the
optically excited state \cite{Fleischhauer2005a}. This
phenomenon lies at the heart of various quantum-optical
control schemes that have been designed for preparing
nonlocal entanglement between spins, quantum communication,
and applying strong optical nonlinearities
\cite{Duan2001,Fleischhauer2005a}.

\begin{figure}
  \includegraphics[width=0.8\linewidth]{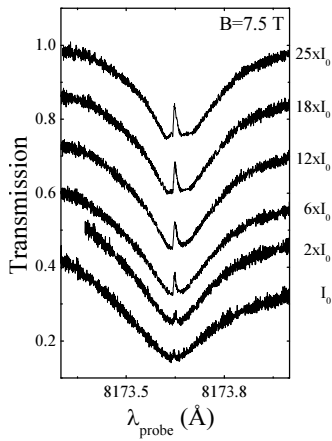}\\
  \caption{Electromagnetically induced transparency
   within the $A^*$ absorption dip,
   induced by a strong control
   field that addresses the $A$ transition.
   Spectra are taken for different intensities $I$
   of the control field,
   with $I_0=0.4~\rm{Wcm^{-2}}$.
   Traces are offset
   vertically for clarity.}\label{fig:EIT}
\end{figure}

EIT can occur with three-level systems as formed by the
states $\mid\uparrow\rangle$, $\mid\downarrow\rangle$ and
$|e\rangle$ (Fig.~\ref{fig:PumpAsSpec}(a)), for which it is
then essential that the two low-energy spin states can have
a long-lived quantum coherence and that one can selectively
address the two optical transitions. An ensemble of these
systems can become transparent for a probe field that
drives one of the transitions (in our case $A^*$) when this
meets the condition for a two-photon Raman resonance with
an applied control field (in our case driving of transition
$A$). Under these conditions the systems are trapped in a
dark state which is in the ideal case $( \Omega_{c}
\mid\uparrow\rangle - \Omega_{p} \mid\downarrow\rangle ) /
\sqrt{|\Omega_{c}|^2 + |\Omega_{p}|^2}$, where $\Omega_{c}$
and $\Omega_{p}$ are the Rabi transition frequencies of the
control and probe field \cite{Fleischhauer2005a,Wang2007}.
Photoluminescence studies on GaAs already showed that
optical control can prepare $D^0$ systems in this dark
state \cite{Fu2005}.


We could demonstrate EIT, and typical results are presented
in Fig.~\ref{fig:EIT}. For these results we fixed the
control laser on resonance with the $A$ transition (V
polarization), while the probe laser is scanned across the
$A^*$ transition (H polarization) and we measure its
transmission. When the control and probe field meet the
condition for two-photon Raman resonance (the difference in
photon energy exactly matches the $D^0$ spin splitting), a
narrow peak with enhanced transmission appears inside the
broader $A^*$ absorption dip. This is the fingerprint of
EIT. In Fig.~\ref{fig:EIT} we present traces for various
intensities of the control field. We observe a wider and
higher EIT peak for stronger driving with the control
field, in agreement with theory for EIT.

EIT relies on quantum coherence between the electron spin
states, and in systems with a very long electron spin
dephasing time $T_2^*$ EIT can fully suppress absorption.
The EIT peak then reaches up to ideal transmission. The EIT
peaks in Fig.~\ref{fig:EIT} are clearly lower, even in the
trace for the strongest control field. From fitting these
EIT traces to the established theory
\cite{Fleischhauer2005a} we derive that the $T_2^*$ value
for our system is about 2~ns, and this compromises the EIT
peak height. This $T_2^*$ value is consistent with earlier
work \cite{Fu2005,Clark2009} that showed that electron spin
dephasing results from hyperfine coupling between each
electron spin and $\sim 10^5$ fluctuating nuclear spins
(the $D^0$ systems have a $\sim 10$~nm Bohr radius). Our
EIT studies also showed weak signatures of dynamical
nuclear polarization (DNP) which confirmed the role of
nuclear spin fluctuations. We anticipate that $T_2^*$ can
be enhanced with controlled DNP effects that suppress the
nuclear spin fluctuations. A longer account of this EIT
study can be found in Ref.~\cite{Sladkov2010}.

\section{Summary}

We presented the realization of a fiber-based confocal
microscope that can be used in a dilution refrigerator
(base temperature well below 20~mK) with high magnetic
field. Faraday rotations in optical materials were
circumvented by using a polarization maintaining fiber and
by having the light propagation in the sample volume in a
direction orthogonal to the applied magnetic field. This
also gives access to performing experiments in Voigt
geometry, which has several advantages. With experiments on
an ensemble of donor-bound electrons in GaAs we confirmed
the ability to focus optical control fields with a small
spot on any desired point of a sample. We also confirmed
that pure linear polarizations can be delivered to the
sample, and that this instrument can perform optical
experiments at milliKelvin temperatures without excessive
heating.

\section{Acknowledgements}

We thank B.~Wolfs, A.~Slachter, M.~Schenkel, J.~Sloot and
A.~Onur for contributions, and J.~De Hosson and
D.~Vainchtein for lending of a spectrometer. Financial
support by the Dutch NWO and FOM, and the German programs
DFG-SPP~1285, BMBF~nanoQUIT and Research school of
Ruhr-Universit\"{a}t Bochum is acknowledged.


\end{document}